\documentclass{Interspeech}

\usepackage{cite}
\usepackage{amsmath,amssymb,amsfonts}
\usepackage{algorithmic}
\usepackage{graphicx}
\usepackage{textcomp}
\usepackage{xcolor}
\usepackage{booktabs}
\usepackage{makecell}
\usepackage{color}
\usepackage{colortbl}
\usepackage{multirow}
\usepackage{multicol}
\usepackage{adjustbox}
\usepackage[flushleft]{threeparttable}
\usepackage{hyperref}
\hypersetup{
    colorlinks=true,
    linkcolor=red,
    anchorcolor=blue,
    citecolor=green
}

\interspeechcameraready

\title{From Sharpness to Better Generalization for Speech Deepfake Detection}

\author[affiliation={}]{Wen}{Huang$^1$}
\author[affiliation={}]{Xuechen}{Liu$^2$}
\author[affiliation={}]{Xin}{Wang$^2$}
\author[affiliation={}]{Junichi}{Yamagishi$^{2\dagger}$}
\author[affiliation={}]{Yanmin}{Qian$^{1\dagger}$}

\affiliation{}{}{$^1$Auditory Cognition and Computational Acoustics Lab}
\affiliation{}{}{MoE Key Lab of Artificial Intelligence, AI Institute}
\affiliation{}{}{School of Computer Science, Shanghai Jiao Tong University, China}
\affiliation{}{}{$^2$National Institute of Informatics, Japan}
\email{holvan@sjtu.edu.cn, jyamagis@nii.ac.jp, yanminqian@sjtu.edu.cn}

\usepackage{comment}

\begin{document}

\maketitle

\begin{abstract}
Generalization remains a critical challenge in speech deepfake detection (SDD). While various approaches aim to improve robustness, generalization is typically assessed through performance metrics like equal error rate without a theoretical framework to explain model performance.
This work investigates sharpness as a theoretical proxy for generalization in SDD. We analyze how sharpness responds to domain shifts and find it increases in unseen conditions, indicating higher model sensitivity. Based on this, we apply Sharpness-Aware Minimization (SAM) to reduce sharpness  explicitly, leading to better and more stable performance across diverse unseen test sets. Furthermore, correlation analysis confirms a statistically significant relationship between sharpness and generalization in most test settings.
These findings suggest that sharpness can serve as a theoretical indicator for generalization in SDD and that sharpness-aware training offers a promising strategy for improving robustness.
\end{abstract}

\renewcommand{\thefootnote}{\dag}
\footnotetext{Corresponding author. Code available at: \href{https://github.com/nii-yamagishilab/SAM-AntiSpoofing}{https://github.com/nii-yamagishilab/SAM-AntiSpoofing}}
\renewcommand{\thefootnote}{\arabic{footnote}}

\section{Introduction}

Speech Deepfake Detection (SDD) aims to differentiate genuine from synthetic speech, which plays a vital role in combating audio-based misinformation, fraud, and security threats. However, generalization remains a major challenge, as models trained on specific datasets often struggle when tested on unseen data, particularly under mismatched conditions such as languages, spoofing attacks, and recording environments~\cite{yamagishi2021asvspoof, muller2022does}. This lack of robustness highlights the need for better generalization strategies to ensure reliable deepfake detection across diverse real-world scenarios.

Various approaches have been proposed to improve generalization in speech deepfake detection, which can be broadly categorized into data-, model-, and feature-level strategies. Data-level methods enhance training diversity through data augmentation~\cite{tak2022rawboost, park2019specaugment} and synthetic spoofed data generation~\cite{wang2023spoofed, wang2024can, zhang2024cpaug} to expose models to a wider range of variations. Model-level approaches improve robustness by leveraging advanced architectures that better capture temporal and spectral information~\cite{jung2022aasist, wu2024spoofing, truong2024temporal} or by utilizing self-supervised learning models such as Wav2Vec 2.0~\cite{baevski2020wav2vec}, XLS-R~\cite{babu2021xlsr}, and WavLM~\cite{chen2022wavlm}, which extract robust and transferable speech representations. Feature-level techniques focus on adjusting the feature space for improved discrimination without modifying model architecture~\cite{zhang2021one, kim2024one, huang2025generalizable}.

Despite these advancements, generalization in speech deepfake detection remains an abstract concept, typically assessed only through performance metrics such as Equal Error Rate (EER) and tandem metrics like the tandem Detection Cost Function (t-DCF)~\cite{kinnunen2020tandem} and tandem EER (t-EER)~\cite{kinnunen2023t}. While these metrics quantify how well a model performs on unseen data, they offer limited theoretical insight into why a model generalizes well or poorly. This lack of interpretability hinders systematic diagnosis of generalization failures and limits principled improvements, which highlights the need for a deeper theoretical understanding to guide more robust systems.

A promising theoretical perspective on generalization comes from sharpness~\cite{keskar2016large}, which quantifies how much a model’s performance changes when its parameters are slightly altered. Intuitively, a sharp model is less robust if small parameter changes cause large performance fluctuations. In contrast, a flatter optimization surface ensures greater stability and better generalization to unseen data.
Many studies suggest that lower sharpness correlates with improved generalization~\cite{keskar2016large, jiang2019fantastic}, motivating the idea of minimizing sharpness to enhance robustness.
To achieve this, Sharpness-Aware Minimization (SAM)~\cite{foret2021sharpnessaware} has been introduced as an optimization strategy that seeks model parameters in regions where the loss remains uniformly low. By jointly minimizing the loss value and sharpness, SAM aims to improve generalization beyond standard training methods. This concept has inspired further research to enhance SAM’s efficiency and effectiveness and has been successfully applied in computer vision and natural language processing. In speech processing, Shim et al.~\cite{shim23c_interspeech} applied SAM to multi-dataset co-training for SDD and demonstrated its potential for improving model robustness on its task.

Nevertheless, the role of sharpness in explaining and quantifying generalization has yet to be explored in SDD. This work aims to bridge this gap by addressing two key research questions:
\begin{itemize}
    \item \textit{Can sharpness serve as a theoretical indicator for generalization in SDD?}
    \item \textit{Can SAM effectively reduce sharpness and enhance generalization performance across diverse SDD datasets?}
\end{itemize}

To explore these questions, we first analyze how sharpness reacts to domain mismatch, examining its behavior across language, attack, channel, and speaker variability. Our findings indicate that sharpness increases in unseen conditions, which showcases its potential to diagnose generalization challenges.
Next, we introduce SAM as an optimization strategy in SDD and evaluate its impact across diverse test sets. Our results show that SAM consistently improves performance, particularly in high-mismatch datasets, by flattening the loss landscape and reducing sharpness.
Finally, we quantify the relationship between sharpness and generalization by computing correlation metrics between sharpness and EER and find a statistically significant correlation in most out-of-distribution datasets.
Overall, these findings suggest that sharpness can serve as a useful theoretical tool for understanding generalization in SDD, while also highlighting the potential of sharpness-aware training to improve robustness.

\section{Sharpness and Domain Mismatch}
\label{sec:sharpness}
Sharpness measures the sensitivity of the loss function to perturbations in the model's parameters, which is often associated with the model's robustness~\cite{foret2021sharpnessaware}. Given this characteristic, we aim to first explore whether sharpness can serve as a diagnostic tool to evaluate how domain mismatch impacts the model's sensitivity and robustness.

To evaluate the sharpness of a model on a specific test set, we adopt the definition provided in~\cite{foret2021sharpnessaware, andriushchenko2022towards}.
Let $S_{test}=\{x_i, y_i\}_{i=1}^n$ denote a test dataset with $n$ samples, where $x_i$ is a sample and $y_i$ is its label, and let $\ell_i(w)$ represent the loss of a classifier parametrized by weights $w$, evaluated at the point $(x_i, y_i)$. Given a maximum allowable perturbation magnitude $\rho$, the sharpness on a subset $S\subseteq S_{test}$ is defined as: 
\begin{equation}
s(w, S) \triangleq \max_{\lVert \epsilon \rVert_2 \leq \rho} \frac{1}{|S|} \sum_{i : (x_i, y_i) \in S} ( \ell_i(w + \epsilon) - \ell_i(w) )
\end{equation}
In practical computations, the sharpness is computed per batch and then averaged over the entire dataset $S_{test}$. This is also referred to as $m$-sharpness, where $m$ represents the batch size.

Since sharpness reflects model sensitivity to domain mismatch, we further investigate its behavior under four mismatch conditions, regarding \emph{language, attack, channel}, and \emph{speaker}. 
For sharpness calculation, we fix the batch size to $m=32$ and the perturbation bound to $\rho=0.05$. We test two models: AASIST~\cite{jung2022aasist} and Wav2vec Base~\cite{baevski2020wav2vec} fine-tuned with linear layers (W2V-Base+Linear). Both models are trained on the ASVspoof 2019 LA training set~\cite{wang2020asvspoof}, which only contains clean English audio. The dataset and training configurations are same as the one provided in in~\ref{sec:setting}. 

\begin{figure}[tb]
    \centering
    \includegraphics[width=0.48\textwidth]{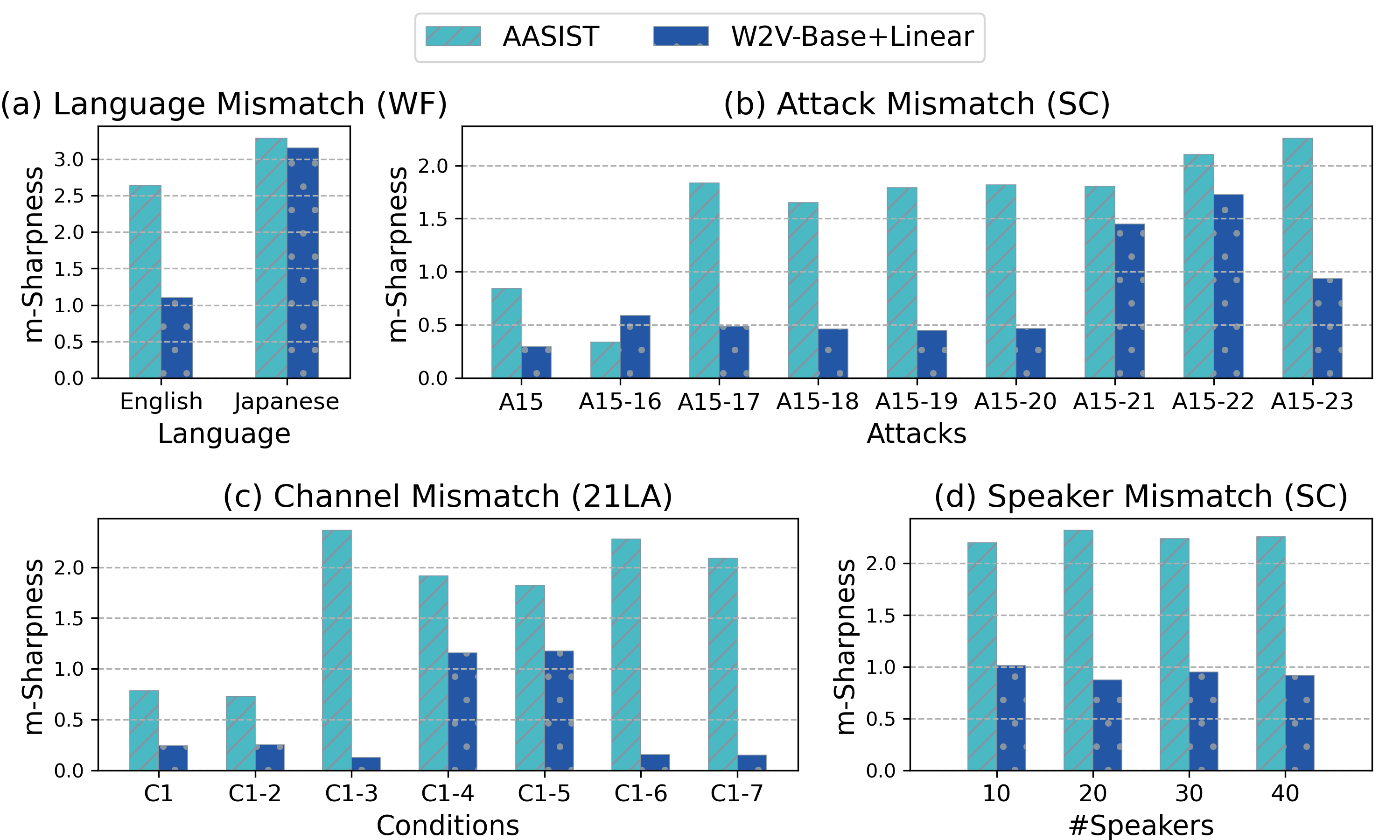}
    \caption{Sharpness values computed for two models, AASIST and W2V-Base+Linear, across different mismatch scenarios. Abbreviations for test sets: WF (WaveFake), SC (SpoofCeleb), 21LA (ASVspoof 2021 LA).}
    \label{fig:mismatch}
\end{figure}

Figure~\ref{fig:mismatch} illustrates the distinct trends observed for each mismatch scenario:


\begin{itemize}
    \item Language: Sharpness is higher for the unseen Japanese subset of WaveFake (WF) than for the English subset, as the training data includes only English.
    \item Attack: Sharpness increases significantly with the number of unseen spoofing attacks in the SpoofCeleb (SC) evaluation set, while the total number of bona fide and spoof samples is kept fixed.
    \item Channel: Sharpness rises with the number of different channel conditions in ASVspoof 2021 LA (21LA), especially for AASIST. In contrast, W2V-Base+Linear shows reduced sensitivity, likely due to robust pretraining and data augmentation.
    \item Speaker: Sharpness remains largely unchanged as the number of speakers increases in SpoofCeleb (SC).
\end{itemize}

These observations show that sharpness naturally reflects sensitivity to mismatch factors, with language, attack, and channel variability increasing sharpness, while speaker variability not significantly affecting it. 
Notably, the W2V-Base+Linear model consistently exhibits lower sharpness than AASIST across all mismatch scenarios. This highlights the advantage of self-supervised learning pretraining in producing robust feature representations.

Based on these findings, we observe a potential relationship between sharpness and generalization. Intuitively, a robust model should exhibit lower sharpness across different test sets, indicating less sensitivity to variations in test data. This raises the question of how sharpness can be minimized to enhance robustness and generalization.

\section{Sharpness-Aware Minimization}

Sharpness-Aware Minimization is proposed in~\cite{foret2021sharpnessaware} to improve model generalization by simultaneously minimizing the loss value and its sharpness. 

Similar to Section~\ref{sec:sharpness}, let $S={(x_i, y_i)}_{i=1}^n$ be a set of labeled data, and let $\ell_i(w)$ denote the loss evaluated at the point $(x_i, y_i)$. For simplicity, we use $L_S(w)=1/|S| \sum_{i=1}^{n}\ell_i(w)$ to denote the accumulated loss for $S$.

To optimize both the loss value and sharpness simultaneously, SAM introduces the following objective function:
\begin{equation}
L_S(w) + \lambda\lVert w\rVert_2^2+[\max_{\lVert\epsilon\rVert_2 \leq \rho} L_S(w + \epsilon) - L_S(w)]
\end{equation}
Here, $L_S(w)$ is the original loss term, and $ \lambda\lVert w\rVert_2^2$ is an L2 regularization term.
The term in square brackets corresponds to the sharpness-aware component, which measures how much the training loss increases when a perturbation $\epsilon$ (bounded by $\lVert \epsilon \rVert_2 \leq \rho$) is added to the model weights. This aligns with the definition of sharpness discussed in Section~\ref{sec:sharpness}. 
After simplification, the SAM objective becomes:
\begin{equation}
\begin{array}{c}
\min_w L_S^{SAM}(w) + \lambda \lVert w\rVert_2^2,\\ 
\text{where}\quad L_S^{SAM}(w) \triangleq \max_{\lVert\epsilon\rVert_2 \leq \rho} L_S(w + \epsilon)
\end{array}
\end{equation}

To efficiently minimize $L_S^{SAM}(w)$, SAM uses an effective approximation by leveraging a first-order expansion of $L_S(w)$ around $w$. The perturbation $\epsilon$ that maximizes $L_S(w + \epsilon)$ is approximated as: 

\begin{equation}
\hat{\epsilon}(w) = \rho \,  \frac{|\nabla_w L_S(w)|}{\lVert\nabla_w L_S(w)\rVert_2}
\end{equation}

Using this approximation, the gradient of $L_S^{SAM}(w)$ can be efficiently computed as:
\begin{equation}
\nabla_w L_S^{SAM}(w) \approx \nabla_w L_S(w) \big|_{w + \hat{\epsilon}(w)}
\end{equation}

This approximation allows SAM to perform sharpness-aware optimization with only a small computational overhead compared with standard gradient descent methods. By encouraging the model to minimize the worst-case loss over a neighborhood of parameter perturbations, SAM achieves flatter loss landscapes and improves the generalization performance of the model.

\section{Experiments and Analysis}
\subsection{Experimental Settings}
\label{sec:setting}

\noindent\textbf{Datasets.}\quad
Systems were trained on the ASVspoof 2019 LA training set~\cite{wang2020asvspoof}, which consists of 25k clean English utterances, including spoofed samples generated using 6 different voice conversion and text-to-speech attacks. The evaluation was conducted on eight unseen datasets representing diverse domain shifts, with performance measured using EER:
\begin{itemize}
    \item ASVspoof 2019 LA evaluation set (19LA)~\cite{wang2020asvspoof} contains 13 spoofing attacks, with 2 being seen and the remaining 11 unseen in the training set.
    \item ASVspoof 2021 LA (21LA)~\cite{yamagishi2021asvspoof} includes 7 channel conditions: 1 clean condition and 6 others affected by different codecs and transmission effects.
    \item ASVspoof 2021 DF (21DF)~\cite{yamagishi2021asvspoof} consists of 600k utterances with over 100 spoofing attacks processed through lossy codecs.
    \item In-The-Wild (ITW)~\cite{muller2022does} contains crowdsourced deepfake and bona fide speech recorded under uncontrolled conditions.
    \item Fake-Or-Real (FOR)~\cite{reimao2019dataset} aggregates deepfake speech from various publicly available sources.
    \item WaveFake (WF)~\cite{frank2021wavefake} includes vocoded speech from LJSpeech (English)~\cite{ljspeech17} and JSUT~\cite{sonobe2017jsut} (Japanese) for language-based evaluations.
    \item ADD 2022 evaluation set (ADD)~\cite{yi2022add} is a Chinese deepfake dataset with multiple spoofing techniques.
    \item SpoofCeleb evaluation set (SC)~\cite{jung2025spoofceleb} contains 40 speakers and 9 spoofing attacks for speaker-dependent evaluations.
\end{itemize}

\noindent\textbf{Models.}\quad
We employed both standard supervised models trained from scratch (non-SSL) and fine-tuned self-supervised learning (SSL) pretrained models. For the non-SSL model, we use AASIST~\cite{jung2022aasist}, an audio anti-spoofing system designed to detect spoofing artifacts.
For SSL models, we utilized Wav2Vec 2.0 \footnote{\href{https:/github.com/facebookresearch/fairseq/tree/main/examples/wav2vec}{github.com/facebookresearch/fairseq/tree/main/examples/wav2vec}} (Base and Large)\cite{baevski2020wav2vec}, a self-supervised framework that learns speech representations from raw audio. Additionally, we incorporated XLS-R\cite{babu2021xlsr}, a multilingual extension of Wav2Vec 2.0 optimized for cross-lingual speech processing. These SSL models were fine-tuned with either linear layers or AASIST~\cite{tak2022automatic} as a backend classifier, denoted as +Linear or +AASIST. 

\noindent\textbf{Training Details.}\quad
The systems were trained across three runs, each using a different random seed. Input speech was randomly chunked into 4-second segments, with RawBoost~\cite{tak2022rawboost} augmentation applied. Training was performed using weighted cross-entropy loss, where the class weights for bona fide and spoof samples were set to 0.9 and 0.1, respectively. The Adam optimizer was used for both standard and SAM training. For non-SSL models, the learning rate was set to 1e-4, following a cosine annealing schedule with a minimum learning rate of 5e-6, and a weight decay of 1e-4. For SSL models, the learning rate was set to 1e-6, with a weight decay of 1e-5. In SAM optimization, the perturbation bound $\rho$ was chosen from $\{0.05, 0.01, 0.005, 0.001\}$, with smaller models assigned larger $\rho$ values.

\subsection{Generalization Improvement with SAM}

\begin{table*}[ht]
\centering
\caption{Performance comparison of different models trained with either Adam or SAM across multiple test sets. The table reports the mean $\pm$ standard deviation EER (\%) over three runs. \textbf{Bold} values indicate the lower value for each model.}
\label{tab:results}
\resizebox{\textwidth}{!}{%
\begin{tabular}{l|c|c|c|c|c|c|c|c|c}
\Xhline{1.2pt}
\textbf{Model} & \textbf{Optimizer} & \textbf{19LA} & \textbf{21LA} & \textbf{21DF} & \textbf{ITW} & \textbf{FOR} & \textbf{WF} & \textbf{ADD} & \textbf{SC} \\ \hline
\multirow{2}{*}{AASIST} & Adam & 1.96 $\pm$ 0.61 & 8.10 $\pm$ 1.33 & 21.92 $\pm$ 2.61 & 34.31 $\pm$ 6.73 & 45.85 $\pm$ \textbf{0.68} & 38.68 $\pm$ 3.02 & 40.48 $\pm$ 3.98 & 45.27 $\pm$ \textbf{3.01} \\
 & \cellcolor{gray!15} SAM & \cellcolor{gray!15} \textbf{1.71} $\pm$ \textbf{0.55} & \cellcolor{gray!15} \textbf{4.62} $\pm$ \textbf{0.81} &  \cellcolor{gray!15} \textbf{19.58} $\pm$ \textbf{1.18} & \cellcolor{gray!15} \textbf{33.34} $\pm$ \textbf{3.07} &  \cellcolor{gray!15} \textbf{33.51} $\pm$ 12.37 & \cellcolor{gray!15} \textbf{33.85} $\pm$ \textbf{2.15} & \cellcolor{gray!15} \textbf{33.84} $\pm$ \textbf{2.80} & \cellcolor{gray!15} \textbf{43.98} $\pm$ 4.71 \\ \hline
\multirow{2}{*}{W2V-Base+Linear} & Adam & 1.78 $\pm$ 0.51 & 5.82 $\pm$ 1.01 & 13.10 $\pm$ 3.00 & 16.85 $\pm$ 2.74 & 12.01 $\pm$ \textbf{2.07} & \textbf{30.34} $\pm$ \textbf{6.71} & 30.47 $\pm$ 9.03 & 34.98 $\pm$ \textbf{1.66} \\
 & \cellcolor{gray!15} SAM & \cellcolor{gray!15} \textbf{1.21} $\pm$ \textbf{0.34} & \cellcolor{gray!15} \textbf{3.39} $\pm$ \textbf{0.89} & \cellcolor{gray!15} \textbf{11.93} $\pm$ \textbf{0.55} & \cellcolor{gray!15} \textbf{13.66} $\pm$ \textbf{1.82} & \cellcolor{gray!15} \textbf{9.57} $\pm$ 2.78 & \cellcolor{gray!15} 36.95 $\pm$ 7.40 & \cellcolor{gray!15} \textbf{24.29} $\pm$ \textbf{3.57} & \cellcolor{gray!15} \textbf{34.87} $\pm$ 2.97 \\ \hline
\multirow{2}{*}{W2V-Base+AASIST} & Adam & 2.81 $\pm$ 0.79 & 4.78 $\pm$ 0.57 & 10.37 $\pm$ 1.19 & 18.29 $\pm$ \textbf{0.47} & 11.11 $\pm$ 1.73 & \textbf{27.79} $\pm$ 3.29 & 36.22 $\pm$ 5.28 & 45.89 $\pm$ 3.56 \\
 & \cellcolor{gray!15} SAM & \cellcolor{gray!15} \textbf{1.32} $\pm$ \textbf{0.38} & \cellcolor{gray!15} \textbf{3.12} $\pm$ \textbf{0.39} & \cellcolor{gray!15} \textbf{10.00} $\pm$ \textbf{0.76} & \cellcolor{gray!15} \textbf{16.21} $\pm$ 2.02 & \textbf{7.92} $\pm$ \textbf{1.37} & \cellcolor{gray!15} 34.29 $\pm$ \textbf{2.88} &  \cellcolor{gray!15} \textbf{28.48} $\pm$ \textbf{3.73} & \cellcolor{gray!15} \textbf{34.83} $\pm$ \textbf{0.46} \\ \hline
\multirow{2}{*}{W2V-Large+Linear} & Adam & 1.37 $\pm$ 0.40 & 3.11 $\pm$ 0.64 & 6.79 $\pm$ 0.38 & 14.96 $\pm$ 1.41 & \textbf{13.24} $\pm$ 2.44 & 23.97 $\pm$ 7.30 & 30.97 $\pm$ 9.91 & 48.35 $\pm$ \textbf{1.40} \\
 & \cellcolor{gray!15} SAM & \cellcolor{gray!15} \textbf{0.88} $\pm$ \textbf{0.10} & \cellcolor{gray!15} \textbf{2.58} $\pm$ \textbf{0.37} & \cellcolor{gray!15} \textbf{6.37} $\pm$ \textbf{0.30} & \cellcolor{gray!15} \textbf{12.22} $\pm$ \textbf{1.41} & \cellcolor{gray!15} 14.65 $\pm$ \textbf{1.79} & \cellcolor{gray!15} \textbf{16.69} $\pm$ \textbf{1.59} & \cellcolor{gray!15} \textbf{37.31} $\pm$ \textbf{6.68} & \cellcolor{gray!15} \textbf{42.89} $\pm$ 5.73 \\ \hline
\multirow{2}{*}{W2V-Large+AASIST} & Adam & 1.22 $\pm$ 0.60 & 4.53 $\pm$ 0.99 & 7.17 $\pm$ \textbf{0.17} & 16.36 $\pm$ \textbf{1.80} & 11.58 $\pm$ 1.81 & 27.89 $\pm$ 2.76 & 33.60 $\pm$ 3.40 & \textbf{39.62} $\pm$ \textbf{4.50} \\
 & \cellcolor{gray!15} SAM & \cellcolor{gray!15} \textbf{1.04} $\pm$ \textbf{0.47} & \cellcolor{gray!15} \textbf{3.60} $\pm$ \textbf{0.16} & \cellcolor{gray!15} \textbf{6.82} $\pm$ 0.32 & \cellcolor{gray!15} \textbf{15.46} $\pm$ 2.09 & \cellcolor{gray!15} \textbf{10.02} $\pm$ \textbf{0.82} & \cellcolor{gray!15} \textbf{25.12} $\pm$ \textbf{1.47} & \cellcolor{gray!15} \textbf{31.41} $\pm$ \textbf{3.28} & \cellcolor{gray!15} 39.67 $\pm$ 5.14 \\ \hline
\multirow{2}{*}{W2V-XLSR+Linear} & Adam & 0.34 $\pm$ 0.06 & \textbf{1.32} $\pm$ \textbf{0.35} & 4.27 $\pm$ \textbf{0.43} & 6.00 $\pm$ \textbf{0.51} & 4.55 $\pm$ 1.19 & 9.87 $\pm$ 3.24 & 22.85 $\pm$ 2.78 & 25.82 $\pm$ \textbf{1.89} \\
 & \cellcolor{gray!15} SAM & \cellcolor{gray!15} \textbf{0.20} $\pm$ \textbf{0.05} & \cellcolor{gray!15} 1.87 $\pm$ 0.39 & \cellcolor{gray!15} \textbf{3.38} $\pm$ 0.47 & \cellcolor{gray!15}  \textbf{5.99} $\pm$ 0.80 & \cellcolor{gray!15} \textbf{3.69} $\pm$ \textbf{0.90} & \cellcolor{gray!15} \textbf{7.66} $\pm$ \textbf{1.24} & \cellcolor{gray!15} \textbf{21.71} $\pm$ \textbf{2.08} & \cellcolor{gray!15} \textbf{25.65} $\pm$ 2.74 \\ \hline
\multirow{2}{*}{W2V-XLSR+AASIST} & Adam & 0.34 $\pm$ 0.13 & 1.85 $\pm$ 0.25 & 3.61 $\pm$ \textbf{0.32} & 6.89 $\pm$ 1.19 & \textbf{4.56} $\pm$ \textbf{0.72} & 16.92 $\pm$ 7.36 & \textbf{19.67} $\pm$ 1.67 & \textbf{27.50} $\pm$ \textbf{1.98} \\
 & \cellcolor{gray!15} SAM & \cellcolor{gray!15} \textbf{0.25} $\pm$ \textbf{0.12} & \cellcolor{gray!15} \textbf{1.71} $\pm$ \textbf{0.27} & \cellcolor{gray!15} \textbf{3.44} $\pm$ 0.54 & \cellcolor{gray!15} \textbf{6.34} $\pm$ \textbf{0.62} & \cellcolor{gray!15} 5.18 $\pm$ 1.48 & \cellcolor{gray!15} \textbf{14.36} $\pm$ \textbf{4.74} & \cellcolor{gray!15} 21.36 $\pm$ \textbf{0.59} & \cellcolor{gray!15} 29.93 $\pm$ 3.30 \\ \Xhline{1.2pt}
\end{tabular}
}
\end{table*}

Table~\ref{tab:results} presents the EER performance of various models trained with either standard Adam or SAM across multiple unseen test sets. The results demonstrate that SAM consistently improves generalization across nearly all models and datasets. The most significant improvements are observed in highly mismatched scenarios, particularly on 21LA, ITW, ADD, and WF, where SAM helps mitigate the impact of unseen conditions. 
Additionally, SAM not only reduces EER but also stabilizes performance, as indicated by the generally lower standard deviations across different models. While some datasets, such as FOR and SC, exhibit relatively higher variance, the reduction in variance for models trained with SAM on most datasets suggests that sharpness minimization leads to more consistent and reliable generalization outcomes.

Beyond optimizer comparisons, the results highlight the advantage of SSL models over non-SSL models. Even when trained with Adam, SSL models generally outperform non-SSL models, indicating that self-supervised pretraining enhances robustness. Notably, SAM further boosts SSL model performance, particularly on datasets with significant domain shifts. This confirms that combining SSL pretraining with sharpness-aware optimization yields stronger generalization.

A more detailed breakdown across models provides deeper insights. AASIST benefits consistently from SAM, with notable improvements on 21LA (4.62 vs. 8.10) and ADD (33.84 vs. 40.48). W2V-Base and W2V-Large also show substantial gains, particularly for W2V-Base+Linear on 21LA (3.39 vs. 5.82) and W2V-Large+Linear on WF (16.69 vs. 23.97). However, for W2V-XLSR, the improvements are less pronounced, suggesting that the model’s pretraining may have already produced a robust feature space and thus reduced the impact of sharpness minimization. These findings demonstrate that while SAM is generally effective, its impact varies depending on the model architecture and dataset characteristics.

\begin{figure}[tb]
    \centering
    \includegraphics[width=0.47\textwidth]{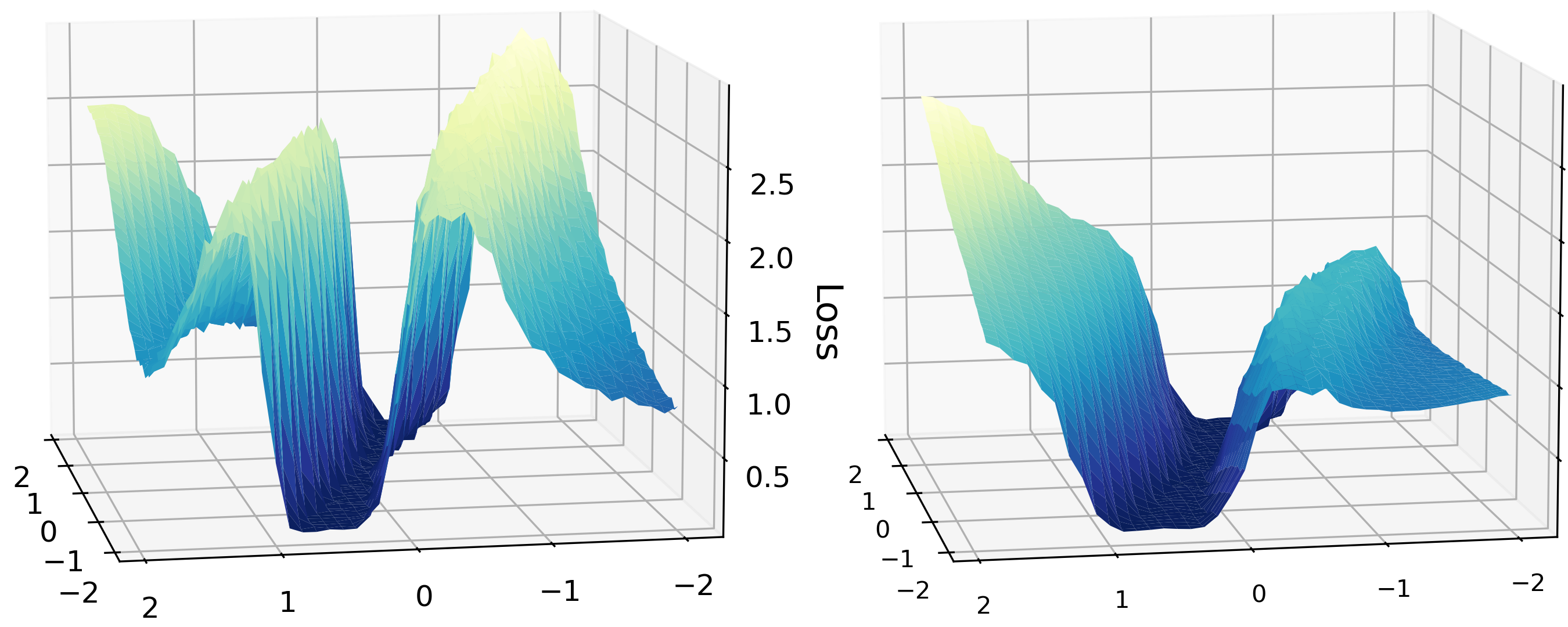}
    \caption{Visualization of the loss landscape for W2V-Base+Linear trained with the Adam optimizer (left) and the SAM optimizer (right). The x-y axes represent scaling factors for two weight perturbation directions, with the original model at (0,0). The z-axis shows the corresponding loss values for the perturbed model.}
    \label{fig:landscape}
\end{figure}

\begin{figure*}[tb]
    \centering
    \includegraphics[width=1\textwidth]{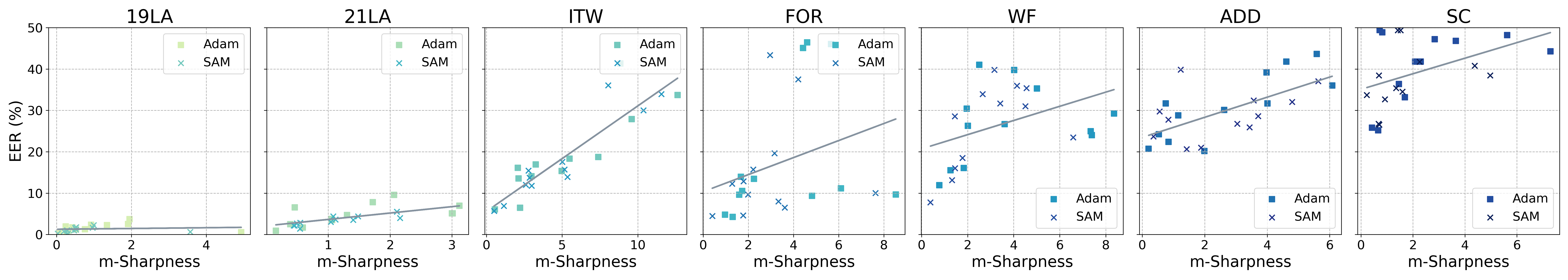}
    \caption{Scatter plots of sharpness (x-axis) and EER (\%) (y-axis) across seven datasets. Each subplot corresponds to one dataset, with square markers for Adam systems and cross markers for SAM systems. A linear regression trend line is included for each group to represent the relationship between the two variables.}
    \label{fig:correlation}
\end{figure*}

To further understand how SAM influences optimization, we visualized the loss landscapes of W2V-Base+Linear trained with Adam and SAM (Figure~\ref{fig:landscape}). 
The visualizations were generated following~\cite{li2018visualizing}\footnote{\href{https://github.com/marcellodebernardi/loss-landscapes}{github.com/marcellodebernardi/loss-landscapes}}. We selected two random directions in the model's weight space, perturbed the trained model along these directions, and evaluated the loss on the 19LA evaluation set. 
The visualization shows that SAM results in a noticeably flatter loss landscape compared to Adam, indicating reduced sensitivity to perturbations. Additionally, the minima appear wider and smoother, which suggests that SAM encourages solutions that are more stable across variations in model parameters. 

\subsection{Sharpness and Generalization}


\begin{table}[b]
\centering
\caption{Correlation coefficients between sharpness and EER performance across different datasets. Statistically significant correlations (p $\leq$ 0.05) are highlighted with a gray background, while highly significant values (p $\leq$ 0.01) are additionally \underline{underlined}.}
\label{tab:correlation}
\resizebox{0.46\textwidth}{!}{
\begin{tabular}{l|c|c|c|c|c|c|c}
\Xhline{1.2pt}
\textbf{Metric} & \textbf{19LA} & \textbf{21LA} & \textbf{ITW} & \textbf{FOR} & \textbf{WF} & \textbf{ADD} & \textbf{SC} \\
\hline
PCC & 0.13 & \cellcolor{gray!15}\underline{0.65} & \cellcolor{gray!15}\underline{0.89} & 0.30 & \cellcolor{gray!15}{0.40} & \cellcolor{gray!15}\underline{0.65} & \cellcolor{gray!15}{0.43} \\
SRCC & \cellcolor{gray!15}\underline{0.53} & \cellcolor{gray!15}\underline{0.77} & \cellcolor{gray!15}\underline{0.87} & \cellcolor{gray!15}{0.40} & \cellcolor{gray!15}\underline{0.52} & \cellcolor{gray!15}\underline{0.62} & \cellcolor{gray!15}\underline{0.49} \\
KTAU & \cellcolor{gray!15}\underline{0.45} & \cellcolor{gray!15}\underline{0.63} & \cellcolor{gray!15}\underline{0.72} & \cellcolor{gray!15}{0.32} & \cellcolor{gray!15}\underline{0.36} & \cellcolor{gray!15}\underline{0.34} & \cellcolor{gray!15}{0.34} \\
\Xhline{1.2pt}
\end{tabular}}
\end{table}

After observing that SAM improves generalization across multiple unseen test sets and results in a flatter loss landscape, we further investigated the relationship between sharpness and generalization via statistical analysis. To do so, we calculated $m$-sharpness using the same process as described in Section~\ref{sec:sharpness} for a total of 24 systems trained with either Adam or SAM. 

The scatter plots in Figure~\ref{fig:correlation} visualize the relationship between sharpness and EER across different datasets, where SAM-trained models generally exhibit lower sharpness across all datasets. Table~\ref{tab:correlation} presents the correlation coefficients, including Pearson’s Correlation Coefficient (PCC), Spearman’s Rank Correlation Coefficient (SRCC), and Kendall’s Tau (KTAU).



While the overall trend supports a positive correlation between sharpness and EER, the strength of this relationship varies across datasets.
ITW, 21LA, and ADD show the strongest correlations across all three metrics, indicating that sharpness predicts generalization performance more effectively in these settings. 
In contrast, FOR, WF, and SC exhibit moderate correlations. This suggests that while sharpness ranking remains meaningful, other factors may also influence generalization. Additionally, SRCC consistently exceeds PCC, which implies that the relationship between sharpness and EER may not always be strictly linear, but the general trend holds. The weakest correlation appears in 19LA, where PCC is low while SRCC and KTAU remain moderate. Since 19LA is an in-domain evaluation set with training and test conditions that are largely similar, sharpness variations may have a more limited impact on performance.

Despite these variations, sharpness exhibits moderate to strong and statistically significant correlations in five out of six out-of-domain datasets, which highlights its effectiveness as an indicator of generalization.

\section{Conclusion}
\vspace{-2mm}
In this work, we examined sharpness as a theoretical proxy for generalization in speech deepfake detection. Our findings show that sharpness increases under domain shifts, indicating model sensitivity to unseen conditions. Our correlation analysis confirmed its positive association with generalization, especially in high-mismatch scenarios. We further demonstrated that sharpness-aware minimization reduces sharpness and generally improves robustness, leading to better and more stable performance across diverse models and datasets. These results highlight sharpness as a useful diagnostic tool and suggest that sharpness-aware training can enhance model resilience to distribution shifts.

\section{Acknowledgements}
This work was conducted during the first author’s internship at NII, Japan. This study is partially supported by JST AIP Acceleration Research (JPMJCR24U3). This work was also supported in part by China NSFC projects under Grants 62122050 and 62071288, in part by Shanghai Municipal Science and Technology Commission Project under Grant 2021SHZDZX0102.

\bibliographystyle{IEEEtran}

\begin{thebibliography}{10}
\providecommand{\url}[1]{#1}
\csname url@samestyle\endcsname
\providecommand{\newblock}{\relax}
\providecommand{\bibinfo}[2]{#2}
\providecommand{\BIBentrySTDinterwordspacing}{\spaceskip=0pt\relax}
\providecommand{\BIBentryALTinterwordstretchfactor}{4}
\providecommand{\BIBentryALTinterwordspacing}{\spaceskip=\fontdimen2\font plus
\BIBentryALTinterwordstretchfactor\fontdimen3\font minus \fontdimen4\font\relax}
\providecommand{\BIBforeignlanguage}[2]{{%
\expandafter\ifx\csname l@#1\endcsname\relax
\typeout{** WARNING: IEEEtran.bst: No hyphenation pattern has been}%
\typeout{** loaded for the language `#1'. Using the pattern for}%
\typeout{** the default language instead.}%
\else
\language=\csname l@#1\endcsname
\fi
#2}}
\providecommand{\BIBdecl}{\relax}
\BIBdecl

\bibitem{yamagishi2021asvspoof}
J.~Yamagishi, X.~Wang, M.~Todisco, M.~Sahidullah, J.~Patino, A.~Nautsch, X.~Liu, K.~A. Lee, T.~Kinnunen, N.~Evans \emph{et~al.}, ``{ASVspoof} 2021: Accelerating progress in spoofed and deepfake speech detection,'' in \emph{ASVspoof 2021 Workshop-Automatic Speaker Verification and Spoofing Coutermeasures Challenge}, 2021.

\bibitem{muller2022does}
N.~M. M{\"u}ller, P.~Czempin, F.~Dieckmann, A.~Froghyar, and K.~B{\"o}ttinger, ``Does audio deepfake detection generalize?'' \emph{Proc. ISCA Interspeech}, 2022.

\bibitem{tak2022rawboost}
H.~Tak, M.~Kamble, J.~Patino, M.~Todisco, and N.~Evans, ``{RawBoost}: A raw data boosting and augmentation method applied to automatic speaker verification anti-spoofing,'' in \emph{Proc. ICASSP}.\hskip 1em plus 0.5em minus 0.4em\relax IEEE, 2022, pp. 6382--6386.

\bibitem{park2019specaugment}
D.~S. Park, W.~Chan, Y.~Zhang, C.-C. Chiu, B.~Zoph, E.~D. Cubuk, and Q.~V. Le, ``{SpecAugment}: A simple data augmentation method for automatic speech recognition,'' \emph{arXiv preprint arXiv:1904.08779}, 2019.

\bibitem{wang2023spoofed}
X.~Wang and J.~Yamagishi, ``Spoofed training data for speech spoofing countermeasure can be efficiently created using neural vocoders,'' in \emph{Proc. ICASSP}.\hskip 1em plus 0.5em minus 0.4em\relax IEEE, 2023, pp. 1--5.

\bibitem{wang2024can}
------, ``Can large-scale vocoded spoofed data improve speech spoofing countermeasure with a self-supervised front end?'' in \emph{Proc. ICASSP}.\hskip 1em plus 0.5em minus 0.4em\relax IEEE, 2024, pp. 10\,311--10\,315.

\bibitem{zhang2024cpaug}
L.~Zhang, K.~A. Lee, L.~Zhang, L.~Wang, and B.~Niu, ``{CPAUG}: Refining copy-paste augmentation for speech anti-spoofing,'' in \emph{ICASSP 2024-2024 IEEE International Conference on Acoustics, Speech and Signal Processing (ICASSP)}.\hskip 1em plus 0.5em minus 0.4em\relax IEEE, 2024, pp. 10\,996--11\,000.

\bibitem{jung2022aasist}
J.-w. Jung, H.-S. Heo, H.~Tak, H.-j. Shim, J.~S. Chung, B.-J. Lee, H.-J. Yu, and N.~Evans, ``{AASIST}: Audio anti-spoofing using integrated spectro-temporal graph attention networks,'' in \emph{Proc. ICASSP}.\hskip 1em plus 0.5em minus 0.4em\relax IEEE, 2022, pp. 6367--6371.

\bibitem{wu2024spoofing}
H.~Wu, W.~Guo, Z.~Zhang, W.~Zhao, S.~Peng, and J.~Zhang, ``Spoofing speech detection by modeling local spectro-temporal and long-term dependency,'' in \emph{Proc. ISCA Interspeech}, 2024, pp. 507--511.

\bibitem{truong2024temporal}
D.-T. Truong, R.~Tao, T.~Nguyen, H.-T. Luong, K.~A. Lee, and E.~S. Chng, ``Temporal-channel modeling in multi-head self-attention for synthetic speech detection,'' in \emph{Proc. ISCA Interspeech}, 2024, pp. 537--541.

\bibitem{baevski2020wav2vec}
A.~Baevski, Y.~Zhou, A.~Mohamed, and M.~Auli, ``wav2vec 2.0: A framework for self-supervised learning of speech representations,'' \emph{Advances in neural information processing systems}, vol.~33, pp. 12\,449--12\,460, 2020.

\bibitem{babu2021xlsr}
A.~Babu, C.~Wang, A.~Tjandra, K.~Lakhotia, Q.~Xu, N.~Goyal, K.~Singh, P.~von Platen, Y.~Saraf, J.~Pino, A.~Baevski, A.~Conneau, and M.~Auli, ``{XLS-R}: Self-supervised cross-lingual speech representation learning at scale,'' \emph{arXiv}, vol. abs/2111.09296, 2021.

\bibitem{chen2022wavlm}
S.~Chen, C.~Wang, Z.~Chen, Y.~Wu, S.~Liu, Z.~Chen, J.~Li, N.~Kanda, T.~Yoshioka, X.~Xiao \emph{et~al.}, ``{WavLM}: Large-scale self-supervised pre-training for full stack speech processing,'' \emph{IEEE Journal of Selected Topics in Signal Processing}, vol.~16, no.~6, pp. 1505--1518, 2022.

\bibitem{zhang2021one}
Y.~Zhang, F.~Jiang, and Z.~Duan, ``One-class learning towards synthetic voice spoofing detection,'' \emph{IEEE Signal Processing Letters}, vol.~28, pp. 937--941, 2021.

\bibitem{kim2024one}
H.~M. Kim, K.~Jang, and H.~Kim, ``One-class learning with adaptive centroid shift for audio deepfake detection,'' in \emph{Proc. ISCA Interspeech}, 2024, pp. 4853--4857.

\bibitem{huang2025generalizable}
W.~Huang, Y.~Gu, Z.~Wang, H.~Zhu, and Y.~Qian, ``Generalizable audio deepfake detection via latent space refinement and augmentation,'' in \emph{Proc. ICASSP}, 2025, pp. 1--5.

\bibitem{kinnunen2020tandem}
T.~Kinnunen, H.~Delgado, N.~Evans, K.~A. Lee, V.~Vestman, A.~Nautsch, M.~Todisco, X.~Wang, M.~Sahidullah, J.~Yamagishi \emph{et~al.}, ``Tandem assessment of spoofing countermeasures and automatic speaker verification: Fundamentals,'' \emph{IEEE Trans. Audio, Speech, Language Process.}, vol.~28, pp. 2195--2210, 2020.

\bibitem{kinnunen2023t}
T.~H. Kinnunen, K.~A. Lee, H.~Tak, N.~Evans, and A.~Nautsch, ``{t-EER}: Parameter-free tandem evaluation of countermeasures and biometric comparators,'' \emph{IEEE Transactions on Pattern Analysis and Machine Intelligence}, 2023.

\bibitem{keskar2016large}
N.~S. Keskar, D.~Mudigere, J.~Nocedal, M.~Smelyanskiy, and P.~T.~P. Tang, ``On large-batch training for deep learning: Generalization gap and sharp minima,'' 2017.

\bibitem{jiang2019fantastic}
Y.~Jiang, B.~Neyshabur, H.~Mobahi, D.~Krishnan, and S.~Bengio, ``Fantastic generalization measures and where to find them,'' 2020.

\bibitem{foret2021sharpnessaware}
P.~Foret, A.~Kleiner, H.~Mobahi, and B.~Neyshabur, ``Sharpness-aware minimization for efficiently improving generalization,'' in \emph{International Conference on Learning Representations}, 2021.

\bibitem{shim23c_interspeech}
H.-j. Shim, J.-w. Jung, and T.~Kinnunen, ``Multi-dataset co-training with sharpness-aware optimization for audio anti-spoofing,'' in \emph{Proc. ISCA Interspeech}, 2023, pp. 3804--3808.

\bibitem{andriushchenko2022towards}
M.~Andriushchenko and N.~Flammarion, ``Towards understanding sharpness-aware minimization,'' in \emph{International Conference on Machine Learning}.\hskip 1em plus 0.5em minus 0.4em\relax PMLR, 2022, pp. 639--668.

\bibitem{wang2020asvspoof}
X.~Wang, J.~Yamagishi, M.~Todisco, H.~Delgado, A.~Nautsch, N.~Evans, M.~Sahidullah, V.~Vestman, T.~Kinnunen, K.~A. Lee \emph{et~al.}, ``{ASVspoof} 2019: A large-scale public database of synthesized, converted and replayed speech,'' \emph{Computer Speech \& Language}, vol.~64, p. 101114, 2020.

\bibitem{reimao2019dataset}
R.~Reimao and V.~Tzerpos, ``{FoR}: A dataset for synthetic speech detection,'' in \emph{2019 International Conference on Speech Technology and Human-Computer Dialogue (SpeD)}.\hskip 1em plus 0.5em minus 0.4em\relax IEEE, 2019, pp. 1--10.

\bibitem{frank2021wavefake}
J.~Frank and L.~Sch{\"o}nherr, ``{WaveFake}: A data set to facilitate audio deepfake detection,'' in \emph{Thirty-fifth Conference on Neural Information Processing Systems Datasets and Benchmarks Track}, 2021.

\bibitem{ljspeech17}
K.~Ito and L.~Johnson, ``The {LJ Speech} dataset,'' \url{https://keithito.com/LJ-Speech-Dataset/}, 2017.

\bibitem{sonobe2017jsut}
R.~Sonobe, S.~Takamichi, and H.~Saruwatari, ``{JSUT} corpus: Free large-scale japanese speech corpus for end-to-end speech synthesis,'' \emph{arXiv preprint arXiv:1711.00354}, 2017.

\bibitem{yi2022add}
J.~Yi, R.~Fu, J.~Tao, S.~Nie, H.~Ma, C.~Wang, T.~Wang, Z.~Tian, Y.~Bai, C.~Fan \emph{et~al.}, ``{ADD} 2022: the first audio deep synthesis detection challenge,'' in \emph{Proc. ICASSP}.\hskip 1em plus 0.5em minus 0.4em\relax IEEE, 2022, pp. 9216--9220.

\bibitem{jung2025spoofceleb}
J.-w. Jung, Y.~Wu, X.~Wang, J.-H. Kim, S.~Maiti, Y.~Matsunaga, H.-j. Shim, J.~Tian, N.~Evans, J.~S. Chung \emph{et~al.}, ``{SpoofCeleb}: Speech deepfake detection and {SASV} in the wild,'' \emph{IEEE Open Journal of Signal Processing}, 2025.

\bibitem{tak2022automatic}
H.~Tak, M.~Todisco, X.~Wang, J.-w. Jung, J.~Yamagishi, and N.~Evans, ``Automatic speaker verification spoofing and deepfake detection using wav2vec 2.0 and data augmentation,'' in \emph{The Speaker and Language Recognition Workshop}, 2022.

\bibitem{li2018visualizing}
H.~Li, Z.~Xu, G.~Taylor, C.~Studer, and T.~Goldstein, ``Visualizing the loss landscape of neural nets,'' \emph{Advances in neural information processing systems}, vol.~31, 2018.

\end{thebibliography}

\end{document}